PLOS COMPUTATIONAL BIOLOGYRESEARCH ARTICLE

# Patterning the insect eye: From stochastic to deterministic mechanisms

**Haleh Ebadi[1]\*, Michael Perry[2], Keith Short[2], Konstantin Klemm[3,4], Claude Desplan[2], Peter F. Stadler[1,5,6], Anita Mehta[5]**

**1** Bioinformatics, Institute for Computer Science, Leipzig University, Leipzig, Germany, **2** Department of Biology, New York University, New York, New York, United States of America, **3** Department of Computer Science, School of Science and Technology, Nazarbayev University, Astana, Republic of Kazakhstan, **4** Instituto de Física Interdisciplinar y Sistemas Complejos, Palma de Mallorca, Spain, **5** Max Planck Institute for Mathematics in the Sciences, Leipzig, Germany, **6** Santa Fe Institute, Santa Fe, New Mexico, United States of America

\* haleh@bioinf.uni-leipzig.de
## Abstract

While most processes in biology are highly deterministic, stochastic mechanisms are sometimes used to increase cellular diversity. In human and Drosophila eyes, photoreceptors sensitive to different wavelengths of light are distributed in stochastic patterns, and one such patterning system has been analyzed in detail in the Drosophila retina. Interestingly, some species in the dipteran family Dolichopodidae (the "long legged" flies, or "Doli") instead exhibit highly orderly deterministic eye patterns. In these species, alternating columns of ommatidia (unit eyes) produce corneal lenses of different colors. Occasional perturbations in some individuals disrupt the regular columns in a way that suggests that patterning occurs via a posterior-to-anterior signaling relay during development, and that specification follows a local, cellular-automaton-like rule. We hypothesize that the regulatory mechanisms that pattern the eye are largely conserved among flies and that the difference between unordered Drosophila and ordered dolichopodid eyes can be explained in terms of relative strengths of signaling interactions rather than a rewiring of the regulatory network itself. We present a simple stochastic model that is capable of explaining both the stochastic Drosophila eye and the striped pattern of Dolichopodidae eyes and thereby characterize the least number of underlying developmental rules necessary to produce both stochastic and deterministic patterns. We show that only small changes to model parameters are needed to also reproduce intermediate, semi-random patterns observed in another Doli species, and quantification of ommatidial distributions in these eyes suggests that their patterning follows similar rules.
## Author summary

A simple model is able to account for a diversity of photoreceptor patterns in different fly species, ranging from highly deterministic to fully random.

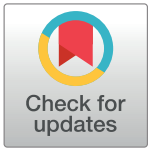

OPEN ACCESS

**Citation:** Ebadi H, Perry M, Short K, Klemm K, Desplan C, Stadler PF, et al. (2018) Patterning the insect eye: From stochastic to deterministic mechanisms. PLoS Comput Biol 14(11): e1006363. https://doi.org/10.1371/journal.pcbi.1006363

**Editor:** Robert Johnston, Johns Hopkins University, UNITED STATES

**Received:** March 23, 2017

**Accepted:** July 16, 2018

**Published:** November 15, 2018
**Copyright:** © 2018 Ebadi et al. This is an open access article distributed under the terms of the Creative Commons Attribution License, which permits unrestricted use, distribution, and reproduction in any medium, provided the original author and source are credited.
**Data Availability Statement:** All relevant data are within the paper and its Supporting Information files.

**Funding:** This work was supported in part by German research foundation STA 850/15-1 to PFS, NIH (grant R01 EY13010) to CD, NIH grant K99 EY027016 to MP Ministerio de Economía, Industria y Competitividad  http://www.mineco.gob.es/ through the Ramón y Cajal program and through project SPASIMM, FIS2016-80067-P (AEI/FEDER, EU) to KK. The funders had no role in study design,
PLOS Computational Biology | https://doi.org/10.1371/journal.pcbi.1006363   November 15, 2018   1 / 15

## Introduction

The development of multicellular animals is highly reproducible, with deterministic and orderly processes generating reliable outcomes. Segment boundaries form in the proper place and cell types are set aside in specific proportions in differentiating tissues. Underlying these seemingly precise developmental outcomes, though, are inherently stochastic transcriptional events, e.g. decisions to express or not express key regulators of cell fate [1, 2]. Varying amounts of activating or repressive input can bias these decisions strongly one way or the other, producing seemingly deterministic on or off outcomes, resulting in distinct boundaries and specific spatial patterns [3, 4]. The distribution of these inputs depends largely on lineage and positional information within an embryo or tissue. In another class of cell fate decisions, stochastic cell-intrinsic mechanisms instead produce a particular probability of taking one fate or another in otherwise equivalent cells [5]. In their own way, these stochastic decisions are highly regulated to take place in specific tissue types and to produce reliable proportions of one cell fate vs. another.

How such probabilistic patterning mechanisms might be switched between stochastic and deterministic is a question to which the tools of statistical physics can meaningfully be applied. An example of stochastic patterning occurs in the Drosophila eye [5, 6], a complex organ whose development has been the subject of great scrutiny [7, 8]. Our interest focuses on the patterning of photoreceptors (PRs) that are involved in color vision: two types of ommatidia express different combinations of color-sensitive photopigments Rhodopsins in their "inner" R7 and R8 PRs, and these two ommatidial types are randomly distributed across the retina [9]. This can be visualized via staining with antibodies against the green-sensitive (Rh6) or blue-sensitive (Rh5) Rhodopsins expressed in R8 PR cells (Fig 1a). A similar stochastic pattern exists in R7 PR cells for two UV Rhodopsins, Rh3 and Rh4 [10]. Stochastic on or off expression of the transcription factor Spineless in the R7 PRs controls ommatidial type, and therefore the overall random pattern [5]. In contrast, another group of flies in the family Dolichopodidae (referred to here as "Doli") have ordered retinal patterning with alternating columns of ommatidia (the individual units of the adult compound eye) that produce two distinct corneal lens colors (Fig 1b). The patterning mechanisms that underlie both differentiation of PR types (e.g. R7 vs. R8) and stochastic patterning across ommatidia have been shown to be largely conserved between Drosophila and butterflies [11, 12]. Considering the apparent similarities between the Drosophila and Doli eye, it is tempting to suggest that the cell fate decisions involved in stochastic vs. non-stochastic patterning may share the same underlying regulatory mechanisms with similar downstream effectors, but which differ in how the expression of few critical upstream regulators is controlled. Thus, it might be possible to predict the rules that underlie the control of stochastic vs. deterministic patterning, and might underlie the evolutionary conversion from one mode of patterning to the other.

In this work, we present a simple mathematical model for such a regulatory mechanism, and compare our results with experimental data from three fly species. Our model is also predictive and applicable to patterns observed in the eyes of other flies; as an example, we present predictions for the eyes of another species of Dolichopodidae that displays intermediate patterns.

The adult eye is a geometrically regular structure composed of hexagonal unit eyes packed into a grid. Patterning begins with the progression of the morphogenetic furrow, a posterior-to-anterior wave of differentiation. Sequential rounds of signaling produce ∼25 highly ordered rows of ∼30 ommatidia each to make up a total of 800 ommatidia per eye [13]. Each ommatidium contains eight PRs and accessory cells: the six "outer PRs" (R1- R6) express a broad-spectrum Rhodopsin, Rh1, and are required for motion and dim light vision. The two

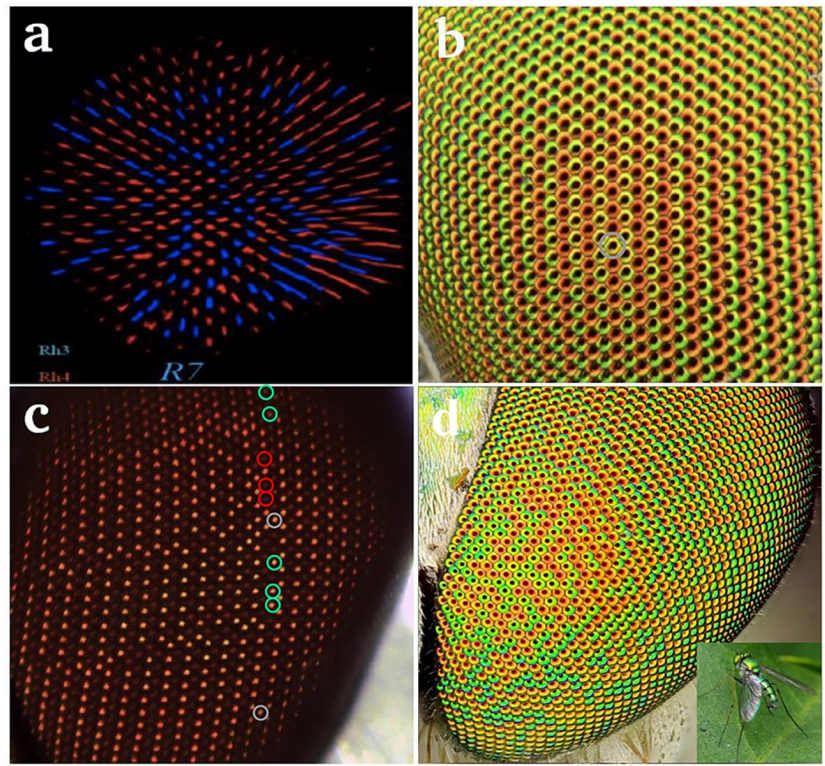

**Fig 1. (a)** Stochastic distribution of UV-sensitive Rhodopsin genes (Rh3 and Rh4) in R7 cells of the Drosophila retina. **(b)** Striped pattern of red alternating with yellow/green ommatidia in the retina in a Condylostylus species fly, family Dolichopodidae. Single "error" circled in grey. **(c)** Image of another Condylostylus individual eye with several errors. Anterior is to the left. In some cases, especially where several errors occur in proximity, the fate of the next most anterior column (to the left) is also modified. Such errors propagate across the eye from posterior to anterior. Note that isolated errors (marked with grey circles) do not propagate. **(d)** Retina of a partially ordered Doli species in the Chrysosoma genus.

https://doi.org/10.1371/journal.pcbi.1006363.g001

"inner PRs" (R7 and R8) each express different Rhodopsins and are used for color discrimination and polarized light vision [8, 14–16]. A detailed mathematical model for much of the process of eye formation has recently been formulated [17]. However, that model does not address the stochastic distribution of color photoreceptors, which is the subject of this paper. There are three main ommatidial subtypes in Drosophila, which are defined by the combination of Rhodopsin photopigments expressed in their R7 and R8 PRs [8].

Two types of so-called **pale** and **yellow** ommatidia, are randomly distributed across the retina in a 35:65 ratio [5, 9]. The **pale** ommatidia express UV-sensitive Rh3 in R7 and blue-sensitive Rh5 in R8 and are used for the discrimination of short-wavelength light [5, 8]. The **yellow** ommatidia express longer UV-sensitive Rh4 in R7 and green-sensitive Rh6 in R8 and are used for the discrimination of longer wavelengths [18] (Fig 1a). A third subtype found in the dorsal rim area (DRA) is used for the detection of the vector of light polarization [19]. The stochastic distribution of **yellow** and **pale** ommatidia in Drosophila is controlled at a single upstream node in the retinal regulatory network by the stochastic expression of the transcription factor Spineless (Ss) in a subset of R7 cells [5]. In Doli, where the patterning is highly ordered, Ss might also be responsible for Rhodopsin expression as the eyes appear to develop in highly similar ways; it also seems likely that many of the interactions in the eye regulatory network are conserved between Doli and Drosophila [11, 12]. The generation of the very different patterns observed might thus be due to changes in the initial expression of Ss. In this paper, we present a simple mathematical model that captures the essence of these ideas, by attributing the diverse patterning in the three fly species (stochastic, ordered and semi-ordered) to a single switching mechanism.

## Model

Early eye development proceeds via a complex set of interactions between cell signaling and changes in target gene expression as new cell types are sequentially recruited [13]. After the progression of the morphogenetic furrow and recruitment of all cell types that will make up the adult retina, cell biological processes begin to shape and structure the ommatidia. The PRs produce microvillae that make up the rhabdomeres, the light-gathering structures. The decision to express Ss (or not) determines the choice of Rhodopsins in the inner photoreceptors R7 and R8, and consequently, the color sensitivity of the ommatidium. In Drosophila, this decision leads to a random distribution of Rhodopsins in **pale** or **yellow** ommatidia [5].

In contrast, Doli eyes show an orderly pattern of alternating columns, such as observed in the genus Condylostylus (Fig 1b). Interestingly, we observed occasional perturbations in patterning in wild-collected Doli (Fig 1c). In some individuals, when multiple errors occur in adjacent or nearly adjacent ommatidia, errors in patterning sometimes propagate in an anterior direction from the initial column containing mistakes (see Fig 1c for an example). In some animals many errors occurred in approximately the same column on the anterior-posterior axis in both retinas, suggesting a developmental cause such as thermal stress at a specific time during the migration of the morphogenetic furrow. Whatever the cause, the subsequent propagation of errors in the direction of the morphogenetic furrow suggests that initially local, cellular-automaton-like rules are at work.

With this locality in mind, we made a key assumption: since the regulatory circuitry is largely conserved between flies and more distant groups such as butterflies, any differences between fly species might be explained in terms of relative strengths of interactions rather than a major rewiring of the regulatory network itself. In this model, we thus assume that there is a single gene product $X$ that is required to activate the switch referred to above, i.e., a deterministic or stochastic "ON" decision to express a specific Rhodopsin can be triggered only in the presence of $X$. We also assume that the coupling strength of the factor X with the switching mechanism that eventually determines the eye color, differs between fly species and can be influenced by changes in the unknown factor $X$ and/or in the switching mechanism itself. Furthermore, the expression of $X$ in an ommatidium should itself depend on the decisions just made in the ommatidia that developed just before it during the progression of the morphogenetic furrow. $X$ is thus a factor that diffuses with the furrow, making local decisions at the point where photoreceptors acquire their identity. In Drosophila, the expression of the transcription factor Spineless leads to the **yellow** state while its absence leads to the **pale** state. In Doli, the color of the cornea (green or red) presumably represents the output of the same circuitry. The conditional probability $P(S|X)$ that $S$ is expressed in a particular ommatidium is assumed to depend—in the general case—on the expression level of $X$. This includes the limiting case $P(S|X) = P(S)$ if the color decision becomes independent of $X$. Since we assume that $X$ is transported with the furrow, we model $X$ as a (weighted) average of the output of $X$ of the preceding ommatidia, i.e., the neighbors of $X$ that have been determined in the previous timestep.

Pattern formation in the fly eye proceeds dynamically column by column: the specification of a particular column occurs following the progress of the furrow across the developing eye. The ommatidia are spatially organized as the furrow moves along [8, 20]. This implies a discrete temporal separation of the columns of ommatidia, which make their color decision at the final step in their formation. The specific parameters clearly depend on the fly species. The Drosophila eye has random ordering with a bias for the **yellow** state. Dolis instead, have ordered and seemingly deterministic retinal patterning, where alternating columns of ommatidia with different corneal colors are found. The color decision is therefore binary. The regular ordering of red and green columns in Doli is occasionally interrupted by mistakes, which suggests that the ordering is also likely non-deterministic in origin. Furthermore, the mistakes that occur in a particular column can propagate over several columns that follow. This suggests that decisions in a column can have at least a local influence on their neighbors. The observed correlation of multiple mistakes in single columns suggests that perturbations can dramatically increase the frequency of mistakes near the location of the morphogenetic furrow at the time of the perturbation. Together, these features suggest a unified theory that contains two ingredients: one, a stochastic choice element, and two, a correlation between the expression of the choice in adjacent columns (which can be set to zero in the case of Drosophila, which remains stochastic). Accordingly, we first define our parameters in the context of the Doli eye and examine its dynamic formation, then discuss the propagation of perturbations. Next, we describe the formation of the Drosophila eye in the context of our model.

## Patterning of the Doli eye

We assume that the choice of a color in each column is essentially complete by the time the morphogenetic furrow moves on to the next column. Thus, we regard fate establishment as instantaneous on the timescale at which the full ordering process occurs in the next column. Consistent with this assumption, Ss expression in Drosophila starts almost immediately after the morphogenetic furrow [21]. In Doli, the two alternate colors, green and red, are denoted by 1 and 0 respectively. In the hexagonal lattice formed by the ommatidia, the color choice of the ommatidium in the $i$th row and $j$th column is defined by the element $a_{ij}$ (which is either 1 or 0) in an $n \times m$ matrix. Two competing effects determine the value of a given element: The first is a default probability of being in the state 1 (green) for every element in a vertical column. The second is that of the subtype correlation between the ommatidium and its nearest neighbors in the preceding column.

## Default probabilities

The default probability $p_j(S|X)$ of all sites in a column j to be green, can be expressed as:

$$p_j(S|X) = \begin{cases} P_0 & X_j \leq X_0 \\ 1 - P_0 & X_j > X_0 \end{cases} \qquad (1)$$

with $P_0$ being a constant. This expresses the fact that, for all values smaller than a threshold $X_0$, the probability for the column to be green is given by $P_0$: this changes to $1 - P_0$ when the threshold is exceeded. High probabilities of being green, as mentioned above, are related to the expression of Spineless. The expression of S leads in its turn to a rapid change in the value of X over column $j + 1$, which can be simply modelled by the following linear equation:

$$X_{j+1} = \gamma - \beta p_j(S|X) \qquad (2)$$

where $\gamma$ and $\beta$ are constants. If $\beta$ is negative, successive columns are anticorrelated, in order to model the case of Doli eyes (Fig 1a), while the positive sign is appropriate for a hypothetical fly that would have a fully homogeneous retina. On the other hand, the constant $\gamma$, which parametrizes the sensitivity of the switch, is strictly positive.

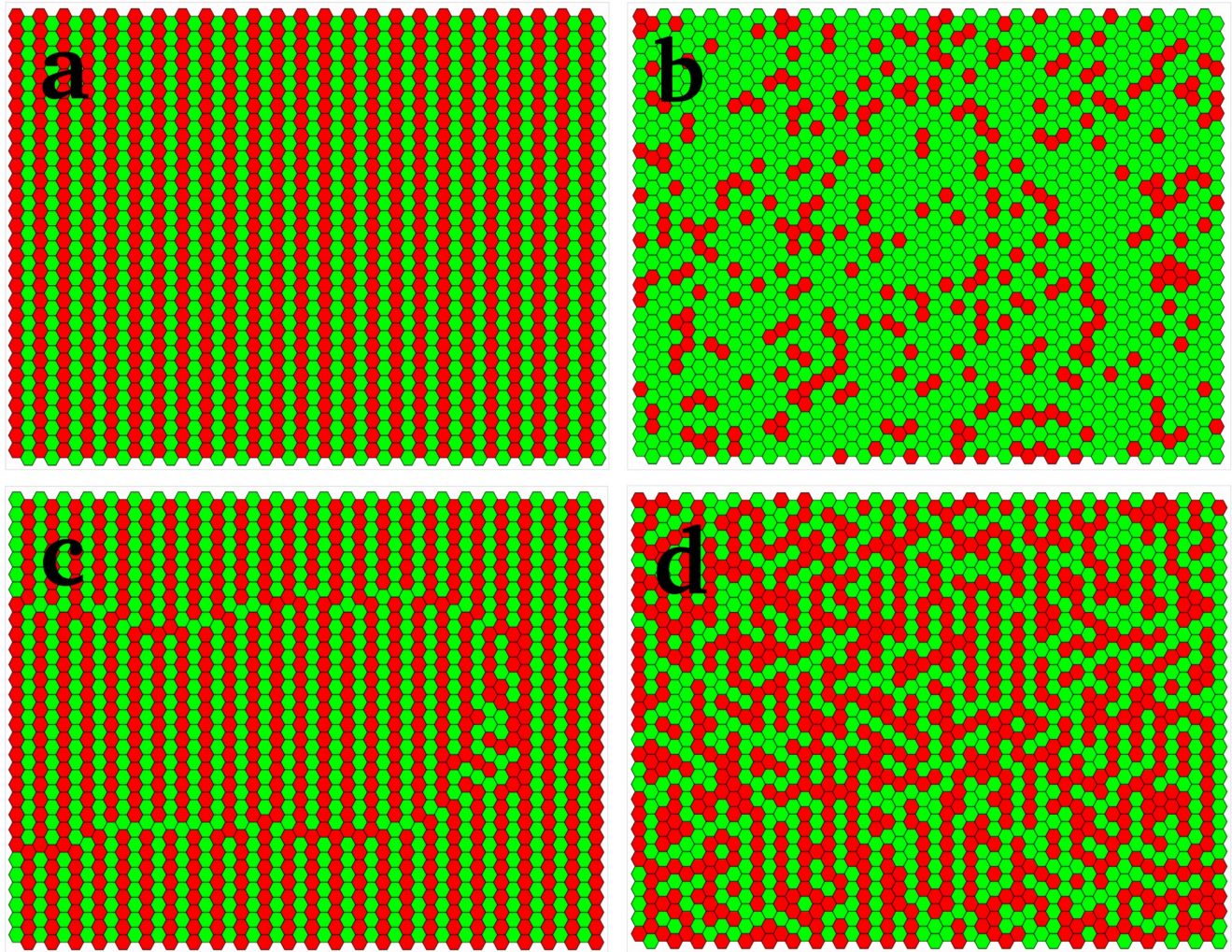

**Fig 2. Simulation results for a model fly eye of size 30 × 50 obtained via the general response function of our model (Eq 6).** The parameter $\alpha$ quantifies the relative importance of stochastic decisions. In: (a) $\alpha = 1$. This models a perfect Doli fly, where there are no mistakes; (b) $\alpha = 0$. This models Drosophila, where the two colors are randomly distributed, with a bias towards the green; (c) $\alpha = 1$. This models a Doli fly a perturbed column (eighth from right). Notice that the perturbations propagate leftward to the end, giving rise to a domain with a discernible boundary. (d) $\alpha = 0.7$. The parameters $\epsilon$ (local speckle correlation coefficient) = 0.95, $P_0^{Doli} = 0.0001$ and $P_0^{Droso} = 0.75$ were chosen with a view to match, at least qualitatively, the patterning of the intermediate, partially organized Chrysosoma sp. shown in Fig 1d.

https://doi.org/10.1371/journal.pcbi.1006363.g002

In order to build an extremely ordered alternating pattern such as that of Doli shown in Fig 1b, we can choose the constant $P_0$ to be very small which guarantees almost uniform color in each column. We illustrate the case for $P_0$ being very small in the following: In Fig 2a, $P_0$ = 0.0001, so that if, at the $j$th column, $X_j$ is greater than the threshold, then the default probability $p_j$ of the column to be green is very close to 1 ($1 - P_0 = 0.9999$, from Eq 1). With suitably chosen constants (e.g. $\beta = 8$, $\gamma = 10$, $X_0 = 5$ as chosen here), Eq 2 implies that the value of $X$ at column $j + 1$, $X_{j+1} \sim 2$, which is less than the threshold $X_0$. In turn, the first line of Eq 1 yields $p_{j+1} = 0.0001$, i.e. column $j + 1$ is red with very high probability. The ordered stripes of red and green are thus built across the eye as depicted in Fig 2a.

For the ordered retina, the negative sign of $\beta$ means that an ordered initial column with an above-threshold value of X, will, from Eq 2, give rise to a constant value of X in all successive columns ($\gamma - \beta(1 - P_0)$), so that a column that is initially green will stay green forever, that is,

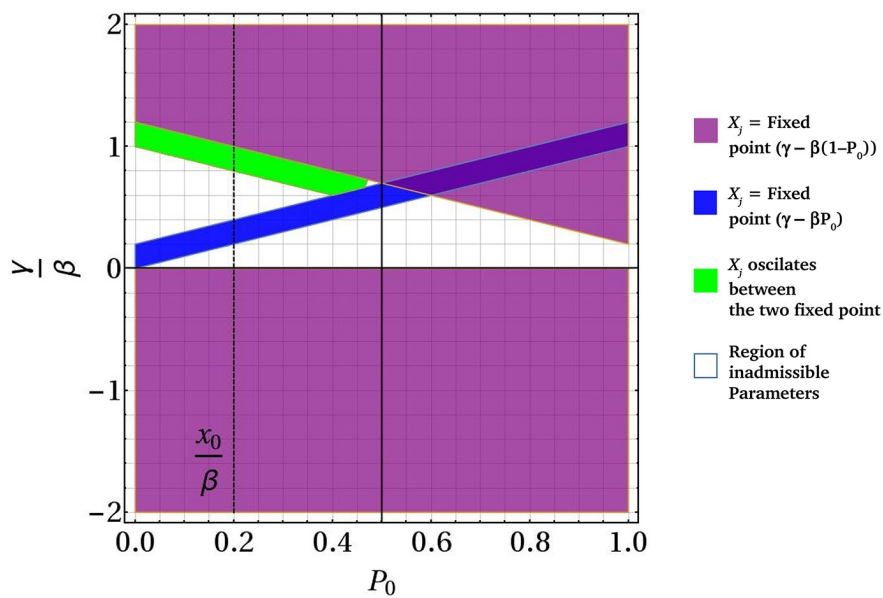

**Fig 3. Plot of the behaviour of $X_j$ based on Eq 2 as a function of the key parameters in the problem with the (scaled) threshold value of $X_0$ shown by the dotted line.** The purple and blue regions correspond to the parameters which lead to fixed values $X_j = \gamma - \beta(1 - P_0)$ and $X_j = \gamma - \beta P_0$ respectively. Such fixed values become stable states of $X_j$, corresponding therefore to the mythical "Uniform Fly". The green region represents the range of parameter values where $X_j$ oscillates between the above fixed points, corresponding therefore to Doli. The white region corresponds to physically inadmissible values where $X_0$ (as a biological factor) takes negative values. In the region where $P_0 > 0.5$, blue, green and purple region overlap; here, $X_j$ goes to a fixed point that depends on its initial value.

https://doi.org/10.1371/journal.pcbi.1006363.g003

all subsequent columns become the same color as the very first one. Some Doli species, such as this example from the Chrysosoma genus (Fig 1d), have partially ordered eyes that are intermediate between Condylostylus and Drosophila retinas. These patterns can be modelled as perturbations of the uniform fly, as seen in Fig 2d.

Fig 3 is an illustration of the quantitative behavior of Eqs 1 and 2 as a function of all the parameters $\beta$, $\gamma$, $P_0$ and $X_0$, with $\gamma$ and $X_0$ scaled by $\beta$ for convenience. Fixing the scaled value $X_0/\beta$ at a sample value of 0.2, we obtain the regions of phase space dominated by Doli-type or uniform retinal pattern. The blue and purple regions correspond to uniform retinas, the green to the Doli retina, while the white region corresponds to physically inadmissible values where $X_0$ (as a biological factor) takes negative values. The overlap of the blue, green and purple region for $P_0 > 0.5$ indicates that the green region or the oscillating $X_j$ will not occur anymore. The value of $X_j$ will be fixed at either $\gamma - \beta(1 - P_0)$ or $\gamma - \beta P_0$ in the overlapped region depending on its initial value.

**Inter-column correlations and propagation of mistakes.** The section above gives the default probabilities in the columns of a Doli retina of a given ommatidium being green: for a perfect column, this is equal to 1 (for a green column) or 0 (for a red column). However, if there are mistakes in the previous column, then the effects of correlation come into play. A green ommatidium in an otherwise fully red column will lead the ommatidia within their correlation neighborhood in the succeeding column also to have mistakes, i.e. be red within a green column. In quantitative terms, the probability is modified by a correction term $l_{ij}$ due to the effect of mistakes, if any. First we define columnar types: a G column is one where most ommatidia are of the 1 type (green), while an R column is where most ommatidia are of the 0 type (red). A mistake is therefore defined as an element that takes a value 1 in an R column or 0 in a G column. Next, we define the correlation neighborhood: in our model, we postulate that correlations decrease exponentially with distance from the target site, so that only the "nearest" mistakes in the previous column would have a significant effect on the ommatidium under consideration. When this correction $l_{ij}$ (see SI for details) is added to the default probability, we have the expression for the full probability of site $ij$ being green:

$$P_{ij}(S, l) = p_j(S|X) + l_{ij} \quad (3)$$

The addition of the correlation factor thus brings about a (positive or negative) correction to the default value of the probability, due to the presence of mistakes in the adjoining column. If there are no mistakes in the earlier column, the default value of the probability is maintained. Fig 2c shows a simulation of a Doli eye with several unique mistakes that have no long-term impact (see also a real Doli eye in Fig 1c). While isolated mistakes such as those shown above usually have little long-distance effect, there are occasions when, possibly due to an accident during development, there are arrays of mistakes extending over several sites in a column. These can then be propagated to adjoining columns over a much longer distance.

Suppose that the defective (normally green) column is $t = 8$, where $p_j = P_{per} = 0.5$: the red ommatidia in Fig 2c correspond to mistakes. These mistakes are in close proximity to each other, leading to a propagation of errors for a considerable distance as the furrow moves onward, sometimes not recovering before the end is reached. By contrast, a single mistake is immediately corrected in the next column, as seen in a picture of a Doli eye (Fig 1b), where the mistake is circled in grey.

### Stochasticity in the Drosophila eye

At the other end of the spectrum from Doli, the Drosophila retina contains ommatidia where R8 cells express either green- or blue-sensitive Rhodopsins that are randomly distributed, with a bias for green-sensitive (**yellow**, 65%) vs. blue-sensitive (**pale**, 35%). This pattern is fully random [5, 22–24]. This implies that the choice in every ommatidium is fully independent of all its neighbors, indicating that there is a total absence of correlations and full stochasticity: no auxiliary equation like Eq 2 is therefore needed. On the other hand, the bias can be accommodated by choosing the default probabilities of the two subtypes to be unequal. The analog of Eq 3 in this case becomes

$$P_{ij}^{Droso}(S, l) = p_j^{Droso}(S|X) \quad (4)$$

where $P_{ij}$ is the probability that the ommatidium $ij$ is green. Fig 2b shows a sample configuration simulated with $P_0 = 0.35$ (see Eq 1), to be compared with an image of the Drosophila eye (Fig 1a).

### Predicting the pattern of other flies

Retinal patterns in other fly species may be conceived as a "mixture" of characteristics of the disorder in Drosophila and Doli or a fully ordered retina, respectively. The simplest way to achieve this from a mathematical perspective is to consider linear combinations of the transition probabilities of the form

$$P_j^\alpha(S|X) = \alpha p_j^{OrderedFly}(S|X) + (1 - \alpha)p_j^{Droso}(S|X), \quad (5)$$

and

$$P_j^\alpha(S|X) = \alpha p_j^{Doli}(S|X) + (1 - \alpha)p_j^{Droso}(S|X), \quad (6)$$

respectively. The parameter $\alpha \in [0, 1]$ quantifies the relative influence of the stochastic,

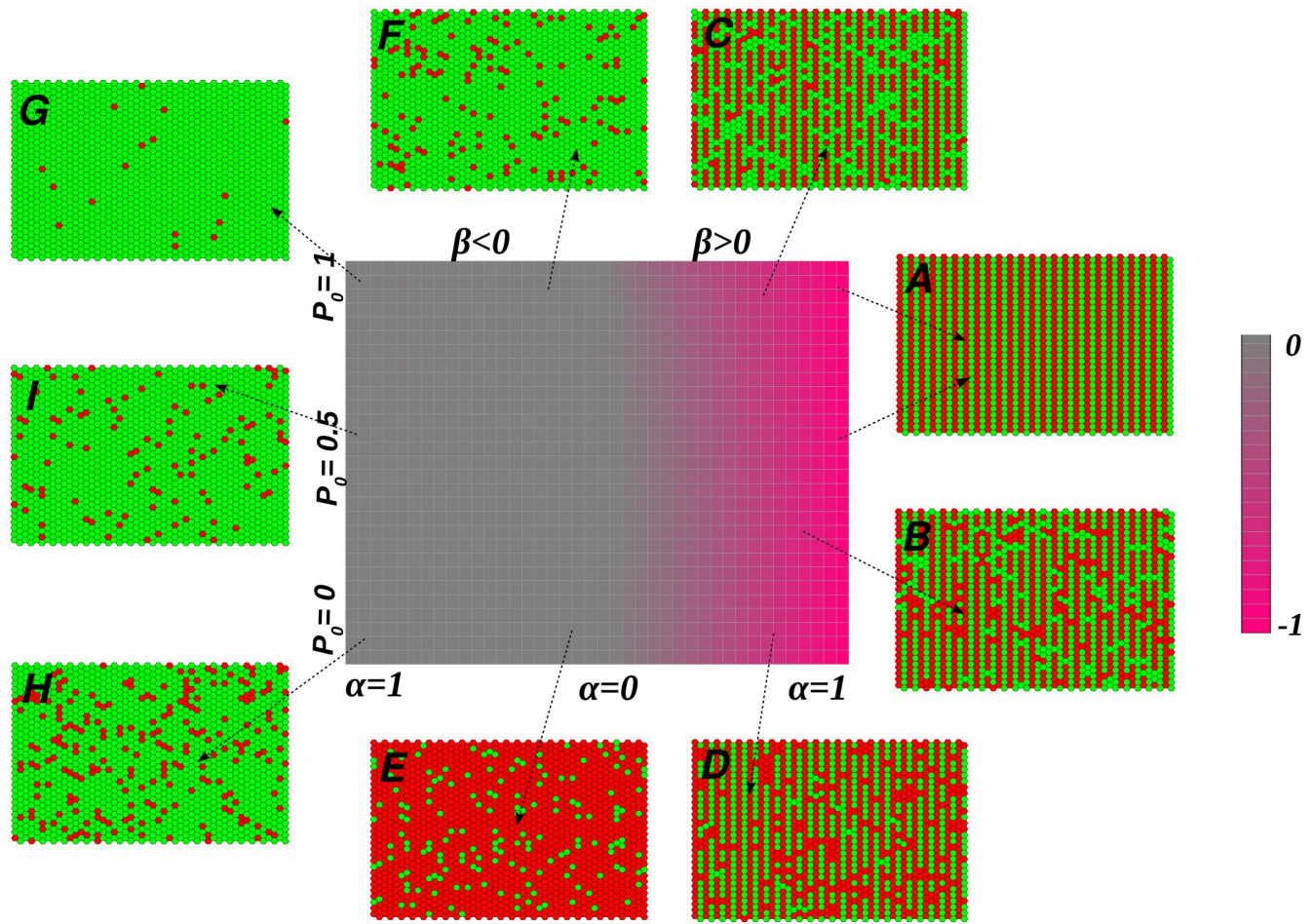

**Fig 4. The mean horizontal correlation coefficient $\langle R^h \rangle$ diagram for a set of situations mimicking various fly eyes according to Eq 9 in S1 Text.** The left half of the main diagram corresponds to $\beta < 0$ (Eq 5) while the right half corresponds to $\beta > 0$ (Eq 6). The horizontal axis shows the value of $\alpha$ which varies from 0 to 1 to the right and left of the y-axis. The vertical axis indicates the value of $P_0^{Droso}$. The chromatic pink indicates $\langle R^h \rangle = -1$, while the gray pixels show $\langle R^h \rangle = 0$. Some examples of the simulated eye configurations are shown (panels (A)-(I)) for the following parameter values: (A) $\beta > 0$, $\alpha = 1$. A value of $\beta > 0$ (Eq 6) indicates the Doli striped region, $\alpha = 1$ makes the second term $((1 - \alpha)p_j^{Droso}(S|X))$ on the right hand side of Eq 6 vanish, so that it becomes independent of $P_0^{Droso}$. We thus have a pure Doli configuration. (B) $\beta > 0$, $\alpha = 0.9$ and $P_0^{Droso} = 0.4$. This configuration deviates slightly from a perfect Doli eye by introducing some randomness via the Drosophila component. (C) $\beta > 0$, $\alpha = 0.65$ and $P_0 = 0.9$. As the value of $\alpha$ is reduced, the effect of the ordered Doli wanes, giving way to patterns that more closely resemble Drosophila. (D) $\beta > 0$, $\alpha = 0.65$ and $P_0 = 0.1$ Here we see Drosophila-like patterns, where the spatial distributions are random, although the ratio of the two colors is fixed. (E) $\alpha$ nearly zero, $P_0^{Droso} = 0.1$ and (F) $\alpha$ nearly zero, $P_0^{Droso} = 0.9$: In both cases, a Drosophila dominant behavior is observed, with the different percentages of reds and greens corresponding to the different values of $P_0^{Droso}$. Note that for $\beta < 0$, i.e. the left half of the phase diagram, the effect of the so-called Uniform Fly increases as $\alpha$ increases at the expense of the Drosophila effect. The extreme case of this is at $\alpha = 1$ when the configurations are a homogeneous green (or red). (G), (H) and (I): $\alpha = 0.9$. with $P_0^{Droso} = 0.9$ (G), $P_0^{Droso} = 0.55$ (I) and $P_0^{Droso} = 0.1$ (H). The dominant behaviour is that of the Uniform Fly, which gives rise to a nearly uniform green(red) color. As $P_0^{Droso}$ increases, red gives way increasingly to green.

https://doi.org/10.1371/journal.pcbi.1006363.g004

*Drosophila*-like mechanism. Eq 5 applies for $\beta < 0$, and Eq 6 for $\beta > 0$. We use this formalism to analyze the eye of a partially ordered fly from the Chrysosoma genus (Fig 1d) where, as for Doli eyes, the pattern is that of corneal colors rather than Rhodopsin expressed in photoreceptors, which are not known in this species. Mathematically, this is a finite $\alpha$ case with $\beta$ negative, corresponding to the first of Eq 5. Our simulations (Fig 2d) agree qualitatively with the image of a real fly shown in Fig 1d; however, a detailed experimental study of its geometrical correlations needs to be performed to get more quantitative agreement.

This formalism allows us to present a generic phase diagram (Fig 4), in which any given fly eye can be defined; the specific location, along with the theory above, enables us to identify the specific mechanisms associated with the morphogenesis, e.g. whether the fly is derived from a perturbation of the Uniform Fly, the hypothetical fly with completely ordered eye colors, or from Doli, and so on. This phase diagram is calculated based on the horizontal and vertical autocorrelations detailed below (see Materials and Methods). Fig 4G corresponds to the case when one of two colors (in this case green) is predominant. The horizontal axis shows the value of $\alpha$ that varies from 0 to 1 symmetrically on each side of the figure; the right side corresponds to positive $\beta$, and the left side to negative $\beta$. The vertical axis gives the value of $P_0^{Droso}$. Panels A-I show sample configurations. Configuration A is pure Doli-like, where $\beta$ is positive and $\alpha = 1$ (the two arrows indicate that the value of $P_0^{Droso}$ is irrelevant, see Eq 6). Configuration B represents a slight deviation from Doli, with $\beta > 0$, $\alpha = 0.9$ and $P_0 = 0.4$. In configuration C, the reduced value of $\alpha = 0.65$ leads to more and more deviations from Doli; here, $P_0 = 0.9$. Configuration D corresponds to parameters $\alpha = 0.65$ and $P_0 = 0.1$; this is a near mirror image of C, because of the flipped value of $P_0$ with respect to C. The $\beta < 0$ region in the negative x-axis is largely disordered. The probability of the so-called Uniform Fly is considered as $p_j^{OrderedFly}(S|X) = 0.99$ so that, e.g. the extreme configuration which occurs at $\alpha = 1$ is 99 percent green. Configurations E and F are drawn based on identical values of $\alpha$ close to zero but for two opposite points on the vertical axis corresponding to $P_0^{Droso} = 0.1$ and $P_0^{Droso} = 0.9$. Therefore, a Drosophila-dominant behavior with different percentages of greens and reds is observed in these panels. As $\alpha$ increases to the left, the effect of the Drosophila term in the first of Eq 6 decreases, so that configurations tend to become more and more a homogeneous green as will be seen in configurations G, I and H which correspond to $\alpha = 0.9$ and $P_0 = 0.9$, 0.55 and 0.1 respectively.

### Parameter inference

We next discuss the significance of our results and seek to relate the patterns we predict theoretically to simulated as well as real data. Since the stochasticity inherent in the model suggests that almost all parameter combinations generate almost all patterns with non-zero probability, the pertinent question to ask is: which parameter combination most likely generates a given pattern?

The aim of this exercise is twofold. First, we show using data simulated by our model, to what extent it is possible to infer the model parameters consistently from the patterns generated by the same model. This provides a baseline for the analysis of real-life images where we strive to infer the most likely parameters for a given image of the fly eye. These parameters then provide the mostly likely explanation for the mechanism that has generated the observed pattern via the phase diagram (Fig 4) given above.

A few technical considerations are in order: The parameters $\beta$ and $\gamma$, cf. Eq (2), themselves are not well-suited for inference because large fractions of the $(\beta, \gamma)$ parameter space generate the same behaviour (see Fig 4). We replace $\beta$ and $\gamma$ by a discrete variable $m$ with only 4 distinct values $m \in \{fr, fg, ar, ag\}$. These represent the combination of two binary variables (i) dynamic mode: fixed point / alternation (ii) initial condition: green probability high / low.

Furthermore, we discretize the values of the parameter $\alpha$ to be multiples of 0.01, i.e. $\alpha \in V(\alpha) := \{0.0, 0.01, 0.02, \ldots, 1.00\}$. We restrict $\epsilon \in V(\epsilon) := \{0.0, 0.1, 0.2, 0.3, 0.4, 0.5\}$ and $P_0 \in V(P_0) := \{0.0, 0.1\}$. This discretization of the inference problem makes it computationally easier. It is justified by the fact that small variations in $P_0$ affect probabilities in a way similar to small variations in $\alpha$. In particular the parameters $\alpha$ and $P_0$ are fully interdependent for a fixed point behavior ($m \in \{fr, fg\}$). In this case, we set $P_0 = 0$. We keep the spatial range parameter at $k = 1.0$ throughout, thus not subjecting it to inference.

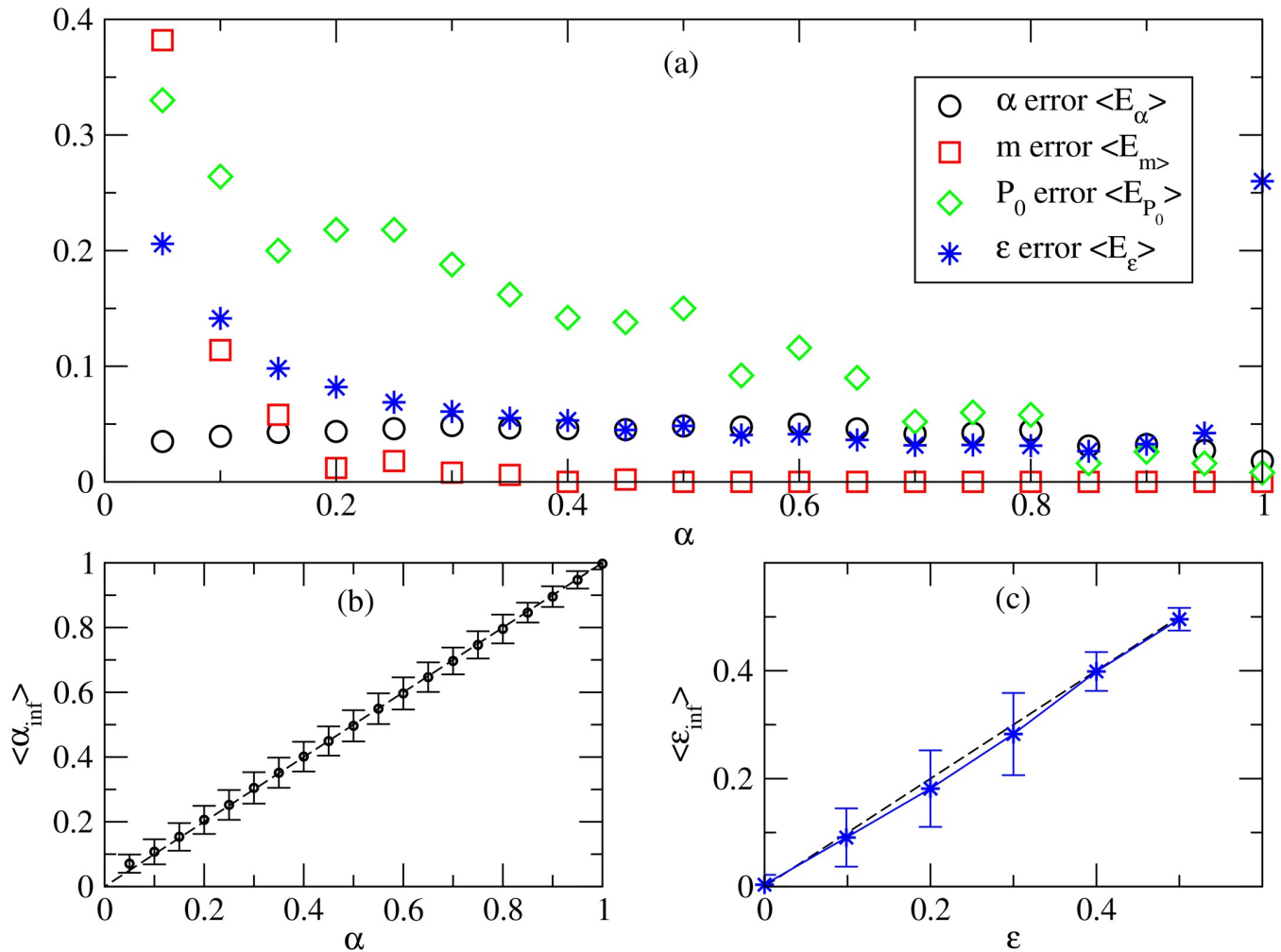

**Fig 5. Inference of model parameters from simulated data.** (a) Average error in parameter inference in dependence of the true value $\alpha$. (b) Average inferred value $\alpha_{inf}$ as a function of true $\alpha$. (c) Average inferred value $\epsilon_{inf}$ as a function of true $\epsilon$. In panels (b) and (c), height of error bars indicates standard deviation; the dashed line is the diagonal, i.e. the identity mapping between true and average inferred values.

https://doi.org/10.1371/journal.pcbi.1006363.g005

We first apply parameter inference to simulated patterns generated by the model itself in order verify that parameters can indeed be inferred consistently. For each realization, we (i) draw parameter values uniformly at random from the set of eligible combinations defined in the preceding paragraph; (ii) generate a pattern by the model with these parameter values, which we call the true values; (iii) find an eligible combination of parameter values that maximizes the probability of the model generating the pattern from step (ii) (maximum likelihood).

Fig 5 shows the results for 10000 independent realizations of the inference using patterns of 50 columns and 30 rows. The mixing parameter $\alpha$ is inferred with good precision over the entire range. Values of the other parameters are also inferred with large precision for sufficiently large $\alpha$. For small $\alpha$, the color assignment is dominated by the fixed probabilities of Drosophila, so that the parameters $m$, $\epsilon$ and $P_0$ have less influence on the pattern. Inference is expected to be more difficult for smaller values of $\alpha$.

As an application of the model to real data, we perform similar parameter inference on photoreceptor distribution data collected from the eyes of individuals in the "intermediate"

Chrysosoma species. See Fig.S1 in S1 Text for details on images and processing of data. Parameters are discretized for the inference; $\alpha$ is in the set {0.01, 0.02, . . ., 0.99}; $P_0 \in$ {0.01, 0.02, 0.04, 0.08, 0.16}; $\epsilon \in$ {0, 0.001, 0.002, . . ., 0.010}; and $k \in$ {1.0, 4.0, 9.0, 16.0, 32.0, . . ., 625.0}.

The results of maximum likelihood parameter inference are as follows:

- At maximum likelihood, we have $m$ = fr throughout, meaning fixed point at red, no oscillation.

- The fraction of red ommatidia is significantly correlated between left and right eyes, rank order correlation (Spearman) $\rho$ = 0.664, P-value (permutation) 0.0041. If one suppresses the part of error propagation in the model (imposing $\epsilon$ = 0), one finds this same correlation also for the $\alpha$ inferred.

- In the full model for which the inference has been done (permitting error propagation), $\alpha$ values are still somewhat correlated between left and right eyes but not significantly, $\rho$ = 0.261, P-value 0.175.

- $\epsilon$ is significantly correlated with $\alpha$ for the same eye; $\rho$ = 0.505, P-value = 0.0023.

- $\epsilon$ values are uncorrelated between left and right eye, $\rho$ = 0.024, P-value = 0.470.

- the inferred spatial scale $\sqrt{k}$ of error propagation is around 15 and thus "almost" the full height of a column. So the scale of error propagation is clearly larger than just nearest neighbor (which would mean $k$ = 1). But it does not span a full column—the errors are somewhat localized vertically.

All in all, the basic behaviour (red fixed point) is found consistently in all eyes. In the other parameters, we have strong fluctuations between samples; in some cases (large $\epsilon$ and large $\alpha$), the stochasticity is described by the large probability of error propagation; in the case of small $\alpha$, the patterning resembles Drosophila. In future studies, stronger support for the model could be obtained by analyzing data from additional species. Ideally, the species could be distinguished by how their parameters fall into disjoint regions.

## Discussion

Statistical physics can be usefully applied to biology when searching for organizational principles [25–27]. The present study is such an example: we present a minimal model that describes patterns observed in the retinas of Drosophila and Dolichopodidae flies. Via a generic phase diagram, we are able to predict parameters that might underlie a range of patterns found on the eyes of various fly species. Conversely, model parameters can be estimated from images of retinal patterns. The essential ingredients of the model are stochasticity and correlations: an ommatidium chooses its color state depending on the competition between its gene-expression-induced probability to be a specific color (e.g. the expression of Spineless) and the influence of the ommatidium in the preceding column along the progression path of the morphogenetic furrow. Intercolumnar correlations greatly outweigh the effect of stochasticity in Doli, while the reverse is the case for Drosophila. This formalism also allows us to probe how mistakes propagate in the retina, for example in Doli. For instance, supposing an ommatidium in a column is a "mistake", i.e. is of the wrong color, our formalism suggests that two scenarios are possible. If the mistake is isolated, it will soon be resolved as the morphogenetic furrow progresses. However, a cluster of mistakes can lead to interesting domain formation, with interfacial roughness as in, say, crystalline systems where dislocations define the area of interfacial disorder [28]. This is in accord with experimental observations [5, 20].

The richness of our formalism, however, goes well beyond the description of known species to the prediction of those that are as yet undiscovered. In terms of simple model parameters, we are able to generate a global phase diagram for flies with binary color choices and allow for important clues to the morphogenetic mechanisms at play. The suggested parameter values for the Chrysosoma species are an example of this power, although we emphasize that in the absence of quantitative experimental data about the molecules involved and variables such as their diffusion constants, we have here aimed at only qualitative agreement. We hope this work will motivate detailed quantitative analysis of experimentally observed patterns, as well as genetic analysis of factors involved in the expression of color in fly eyes, so that our predictions can be put to the test. Once our model is adequately tested for two-color retinae, we will be able to extend our analysis to flies and other insects with more complex color patterning.

## Supporting information

**S1 Text. The technical details of the model and formulations.**
(PDF)

## Acknowledgments

HE would like to thank Ulf-Dietrich Braumann for his helpful guides for visualization of the results.

## Author Contributions

**Conceptualization:** Anita Mehta.

**Data curation:** Haleh Ebadi, Michael Perry, Keith Short, Claude Desplan.

**Formal analysis:** Haleh Ebadi, Konstantin Klemm, Claude Desplan, Peter F. Stadler.

**Funding acquisition:** Michael Perry, Claude Desplan, Peter F. Stadler.

**Investigation:** Haleh Ebadi, Michael Perry, Peter F. Stadler.

**Methodology:** Haleh Ebadi, Peter F. Stadler, Anita Mehta.

**Resources:** Michael Perry, Claude Desplan, Peter F. Stadler.

**Software:** Haleh Ebadi, Konstantin Klemm.

**Supervision:** Konstantin Klemm, Claude Desplan, Peter F. Stadler, Anita Mehta.

**Validation:** Haleh Ebadi.

**Writing – original draft:** Haleh Ebadi, Anita Mehta.

**Writing – review & editing:** Michael Perry, Konstantin Klemm, Claude Desplan, Peter F. Stadler.